\documentclass[sigconf]{acmart} % ,anonymous,review
\usepackage{threeparttable}
\usepackage[utf8]{inputenc} % allow utf-8 input

\usepackage{booktabs}       % professional-quality tables
\usepackage{nicefrac}       % compact symbols for 1/2, etc.
\usepackage{microtype}      % microtypography
\usepackage{xcolor}         % colors
\usepackage{svg}
\usepackage{soul}
\usepackage{url}
\usepackage[font={small,bf}]{caption}
\usepackage{graphicx}
\usepackage{amsmath}
\usepackage{amsthm}
\usepackage{algorithm}
\usepackage[normalem]{ulem}
\usepackage{algorithmic}
\usepackage{multirow}
\usepackage{subfigure}
\usepackage{lipsum}
\usepackage{bm}
\usepackage{color}
\usepackage{mathtools}

\usepackage{tablefootnote}
\usepackage{enumitem}
\usepackage{adjustbox}
\usepackage{tcolorbox}
\setlist[itemize]{leftmargin=*}

\theoremstyle{plain}

\theoremstyle{definition}

\theoremstyle{remark}

\setcopyright{acmlicensed}
\copyrightyear{2018}
\acmYear{2018}
\acmDOI{XXXXXXX.XXXXXXX}

\acmConference[Conference acronym 'XX]{Make sure to enter the correct
  conference title from your rights confirmation emai}{June 03--05,
  2018}{Woodstock, NY}

\acmISBN{978-1-4503-XXXX-X/18/06}

\begin{document}

\title{Crocodile: Cross Experts Covariance for Disentangled Learning in Multi-Domain Recommendation}

\author{Zhutian Lin$^{1}$, Junwei Pan$^{2}$, Haibin Yu$^2$, Xi Xiao$^1$*, Ximei Wang$^2$\\
 Zhixiang Feng$^2$,  Shifeng Wen$^2$, Shudong Huang$^2$, Dapeng Liu$^2$, Lei Xiao$^2$}
 % Zhutian Lin, Junwei Pan, Haibin Yu, Xi Xiao, Ximei Wang, Zhixiang Feng, Shifeng Wen, Shudong Huang, Dapeng Liu and Lei Xiao
\affiliation{%
    \institution{$^1$ Shenzhen International Graduate School, Tsinghua University \,\,
    $^2$ Tencent Inc.
    \country{}
}}

\email{linzt22@mails.tsinghua.edu.cn,
jonaspan@tencent.com}

\renewcommand{\shortauthors}{Z. Lin and J. Pan et. al.}

\begin{abstract}

Multi-domain learning (MDL) has become a prominent topic in enhancing the quality of personalized services. 
It's critical to learn commonalities between domains and preserve the distinct characteristics of each domain.
However, this leads to a challenging dilemma in MDL. 
On the one hand, a model needs to leverage domain-aware modules such as experts or embeddings to preserve each domain's distinctiveness.
On the other hand, real-world datasets often exhibit long-tailed distributions across domains, where some domains may lack sufficient samples to effectively train their specific modules. Unfortunately, nearly all existing work falls short of resolving this dilemma.
To this end, we propose a novel Cross-experts Covariance Loss for Disentangled Learning model (Crocodile), which employs \emph{multiple embedding tables} to make the model domain-aware at the embeddings which consist most parameters in the model, and a \emph{covariance loss} upon these embeddings to disentangle them, enabling the model to capture diverse user interests among domains.
Empirical analysis demonstrates that our method successfully addresses both challenges and outperforms all state-of-the-art methods on public datasets. 
During online A/B testing in Tencent's advertising platform, Crocodile achieves 0.72\% CTR lift and 0.73\% GMV lift on a primary advertising scenario.

\end{abstract}


\maketitle

\section{Introduction}
\label{sec: intro}

\begin{figure}[!ht]
\centering
\subfigure[Dillema of Preserving Distinctiveness v.s. Sufficient Parameters Learning] {\includegraphics[width=.22\textwidth]{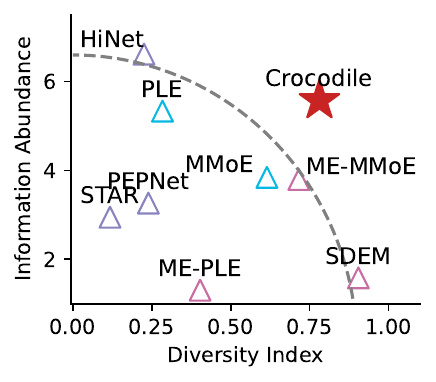}\label{fig: introb}}
\subfigure[gAUC in Kuairand1k and AliCCP] 
{\includegraphics[width=.23\textwidth]{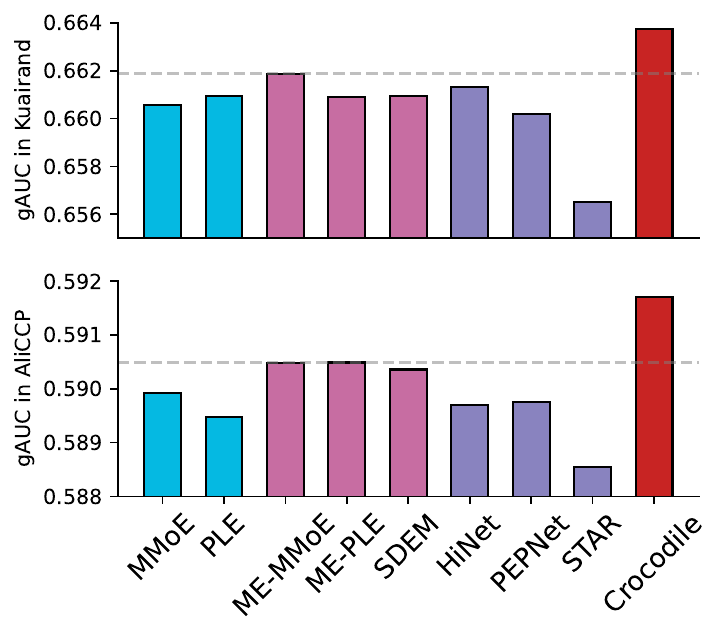}\label{fig: introa}}
\caption{Our proposed Crocodile successfully resolves the dilemma of preserving domain distinctiveness (measured by Diversity Index) v.s. sufficient parameters learning (measured by Information Abundance), and achieves the best performance measured by gAUC.}
\label{fig: intro}
\vspace{-16pt}
\end{figure}

Online recommendation systems~\cite{broder2008computational} are now essential tools for businesses to effectively reach their target audiences, servicing billions of internet users daily. 
Empowered by advanced algorithms and vast traffic, these systems play a crucial role in modeling user experiences and generating revenue for advertisers. 
However, traditional methods still face substantial challenges due to the dynamic nature of online platforms, which are marked by diverse user preferences across domains.
To address these challenges, researchers have increasingly turned to multi-domain learning (MDL)~\cite{chang2023pepnet, zhou2023hinet,sheng2021one,yang2022adasparse,li2023adl} as a promising approach to enhance the performance by involving information from multiple related domains.

However, MDL faces conflicts of interest among users across different domains, which limits the potential for further exploitation of multi-domain information. 
Consequently, existing MDL models attempt to maintain domain distinctiveness through the implementation of domain-aware components along the upside-down trend, \textit{i.e.}, from Tower to Embedding.
Specifically, Shared-Bottom~\cite{SharedBottom1997} and STAR~\cite{sheng2021one} focuses on making the tower domain-specific, while HiNet~\cite{zhou2023hinet}, M-scan~\cite{zhu2024m} and HierRec~\cite{gao2024hierrec} shift the focus to the expert level.
% \footnote{One can view expert to be domain-specific transformed embedding}. 
PEPNet~\cite{chang2023pepnet} follows this trend further by emphasizing domain-specific transformed embeddings.
However, due to the domination of embeddings regarding the number of parameters in recommendation~\cite{pan2024ad}, these experts or transformation networks have limited capacities, making their ability to preserve each domain's distinctiveness in doubt (refer to \emph{Finding 1} in Sec.~\ref{subsec:preserve_distinctiveness}).

Meanwhile, recent work in recommendation shows the efficacy of learning multiple separate embeddings for each feature.
In particular, the Multi-Embedding paradigm~\cite{ME2023, pan2024ad} was proposed in single-task learning for feature interaction models, aiming to resolve the dimensional collapse of embeddings caused by the \emph{Interaction-Collapse theory}~\cite{ME2023}.
Regarding multi-task learning, a variant of multiple embeddings, namely the Shared and Task-specific EMbedding (STEM)~\cite{su2023stem} was proposed to retain each task's distinct characteristics via task-specific embeddings.

Following these works, we try incorporating Shared and Domain-specific EMbeddings (SDEM) in multi-domain learning.
However, these models are still not performing well.
After careful analysis, we found that the embeddings are insufficiently learned, especially for those small domains (refer to \emph{Finding 2} in Sec.~\ref{subsec:sufficient_learning}).
These small domains have much fewer training samples, making it challenging to learn independent embedding tables for them.

In this paper, we propose a novel Diverse Multi-Embeddings paradigm, which adopts multiple shared embeddings with a novel disentanglement loss.
Every embedding table is shared and hence optimized by all domains, overcoming the above-mentioned challenge of embedding insufficient learning, while the disentanglement loss promotes the diversity of these embeddings to preserve the underlying characteristics of all domains, as shown in Fig.~\ref{fig: introb}.
Upon these multiple disentangled embeddings, a novel Prior Informed element-wise Gating (PeG) mechanism was proposed to route between these embeddings and domain-specific classification towers.
We name our method Crocodile, standing for \textbf{Cro}ss-experts \textbf{Co}variance loss for \textbf{Di}sentangled \textbf{Le}arning.

Empirical analysis conducted on two public datasets demonstrates that Crocodile achieves state-of-the-art results, as briefly shown in Fig.~\ref{fig: introa}. Further analysis reveals that Crocodile effectively mitigates embedding collapse while simultaneously learning diverse representations. Ablation experiments confirm the positive contributions of each proposed component to the overall performance. Online deployment on Tencent's advertising platform shows a significant CTR and GMV lift. In summary, our contributions can be summarized as follows:

\begin{itemize}
    \item We discover a dilemma between learning domain-specific characteristics and sufficient learning of parameters of existing multi-domain learning methods.
    \item We propose the Crocodile approach, which employs a Disentangled Multi-Embedding paradigm and a Prior Informed element-wise Gating (PeG) mechanism. 
    \item Empirical evidence on two public datasets as well as online A/B testing on Tencent's advertising platform demonstrate that Crocodile achieves state-of-the-art performance and resolves the aforementioned dilemma.
\end{itemize}
\vspace{-10pt}

\section{On the Challenges of Multi-Domain Recommendation}
\label{sec:challenge}

In this section, we study to what extent existing multi-domain recommendation models resolve the following two challenges: \emph{preserving each domain's distinctiveness} and \emph{sufficient learning of embeddings}.
Besides existing works, inspired by the success of multi-embedding~\cite{su2023stem, ME2023}, we developed two new baselines: 
a Shared and Domain-specific EMbeddings (SDEM) approach, which corresponds to the STEM in MTL,
a Multi-Embedding PLE for multi-domain learning, ME-PLE in short, which corresponds to the ME-PLE in multi-task learning,
and a Multi-Embedding MMoE for multi-domain learning, ME-MMoE in short.
In the following, we evaluate the ability of existing and proposed methods to preserve each domain's distinctiveness and the learning sufficiency of parameters.

\subsection{Preserving Distinctiveness}
\label{subsec:preserve_distinctiveness}
Intuitively, the ability to capture various user interests across domains can be directly characterized by the diversity of representations. 
When facing conflict preference across domains, a set of diverse representations is enabled to encode both prefer/non-prefer semantics and thus provide distinct vectors to towers of those conflict domains. 
To this end, we propose a novel Diversity Index (DI) metric to quantify the diversity among the outputs of experts. 

We select a subset of samples where there is a diverse interest across domains, quantified by the output norm of domain-specific experts.
We denote the set of samples satisfying such conditions as $\mathcal{D}_{(i,j)}$. Formally,

\begin{equation}
    \mathcal{D}_{(i,j)} = \{ k | \|\bm{O}\|^{i}_\text{SD}(x_k) \geq \tau_{t} \cap \|\bm{O}\|^{j}_\text{SD}(x_k) \leq \tau_{b}\},
\end{equation}
where $\tau_{t}$ and $\tau_{b}$ are two specified norm thresholds.

Then, for those models where no experts are bound to any domains such as MMoE or STAR, we simply define the Diversity Index by whether there is one expert that learns a strong interest whilst another one learns a weak interest.
% The interest is measured by the norm of the expert's output.
That is, we just need to count the ratio of samples that have both a strong expert output norm and a weak one:

\begin{equation}
    \frac{\sum\limits_{k\in \mathcal{D}_{(i,j)}} \mathbb{I} [\exists p, q, (\|\mathbf{O}^{p}(x_k)\|_2 \geq \tau_{t}) \cap (\| \mathbf{O}^{q}(x_k) \|_2 \leq \tau_{b})]}{|\mathcal{D}_{(i,j)}|},
\end{equation}

where $\mathbb{I}[\text{\textit{condition}}]$ is the indication function, which is assigned to 1 if the condition is true; otherwise, 0. $p, q$ denote the index of experts, $\mathbf{O}^p(x_k)$ means the output of $p$-th expert of sample $x_k$.

For those models with domain-specific experts, such as PLE, SDEM, and ME-PLE, we can quantify the Diversity Index by counting the ratio of samples that have a large expert output norm on domain $i$'s specific expert while a small norm on $j$'s expert:

\begin{equation}
    \frac{\sum\limits_{k\in \mathcal{D}_{(i,j)}} \mathbb{I} [(\|\mathbf{O}^i(x_k)\|_2 \geq \tau_{t}) \cap (\|\mathbf{O}^j(x_k)\|_2 \leq \tau_{b})]}{|\mathcal{D}_{(i,j)}|}.
\end{equation}

After we compute the diversity measurement $DI_{(i,j)}$ for all domain pairs $(i, j)$, we then take an average of them, getting the overall Diversity Index: $DI = \text{avg} (DI_{(i,j)})$.
We then calculate the diversity index of several existing multi-domain recommendation models and those proposed models with multiple embeddings, as illustrated in Fig.~\ref{fig: introb}.
We have the following findings.

First, existing methods, in general, learn similar interests across domains, as indicated by the low diversity index. 
For example, the diversity index of HiNET, STAR, PEPNet, and PLE are less than 0.3.
That means, for those samples that have shown strong domain-wise interest distinctiveness, these models can capture such distinctiveness for less than 30\% of them.
MMoE performs the best, with a diversity index of 0.615, but it's still not a satisfactory result.
Second, those multi-embedding models have higher diversity indices, especially compared to their single-embedding correspondence.
ME-PLE gets a diversity index 0.403, compared to 0.284 of its single-embedding variant PLE; ME-MMoE gets a diversity index 0.716, compared to 0.615 of its single-embedding variant MMoE.
SDEM gets the highest diversity index among models, \textit{i.e.}, 0.904.
We conclude with the following finding:

\begin{tcolorbox}[colback=blue!2!white,leftrule=2.5mm,size=title]
    \emph{%
        Finding 1. Existing single-embedding recommendation models for multi-domain learning fail to preserve domain distinctiveness.
        When employing multiple embeddings, especially domain-specific embeddings, models can preserve domain distinctiveness more decently.
        }
\end{tcolorbox}

\subsection{Sufficient Parameter Learning}
\label{subsec:sufficient_learning}

We found that the proposed ME-PLE and SDEM models perform poorly, particularly in domains with less data.
For instance, in domain S6 of the Kuairand1k dataset~\cite{gao2022kuairand}, ME-PLE and SDEM show -0.4\% and -0.1\% lower AUC than the single domain method.

We investigated the learned parameters, focusing on the embeddings.
we assess the quality of embedding learning, particularly focusing on dimensional collapse testing, by analyzing the distribution of singular values and employing the aggregation metric Information Abundance (IA)~\cite{guo2023embedding}.
It is defined as the sum of singular values divided by the largest one, \textit{i.e.}, $ ||\mathbf{\sigma}||_1/||\mathbf{\sigma}||_{\infty}$. 

A large IA indicates a balanced distribution of singular values and a robust embedding, while a small IA indicates an imbalanced distribution and dimensional collapse~\cite{guo2023embedding,chi2022representation}.

\begin{figure}[h!]
    \centering
    \subfigure[$log(IA)$ of Experts] {\includegraphics[width=.2\textwidth]{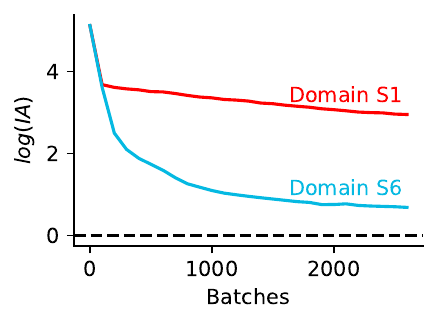}\label{fig: IA_a}}
    \subfigure[$log(IA)$ of Item Embeddings] {\includegraphics[width=.21\textwidth]{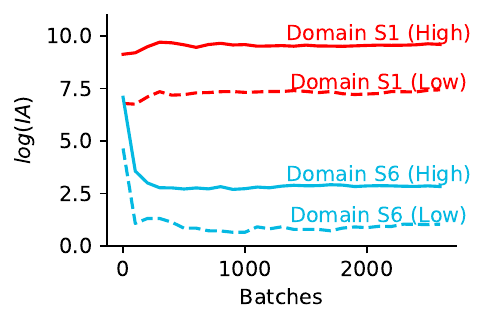}\label{fig: IA_b}}
    \caption{Information Abundance ($log(IA)$) dynamics of SDEM's bottom layer of experts and item embeddings on Kuairand1k. High and Low denote the items with the highest and lowest frequencies.
    }
    \vspace{-10pt}
    \label{fig: ia}
\end{figure}

Taking SDEM as an example, we analyzed its training IA dynamics of experts and embeddings in Fig.~\ref{fig: ia}.
In Kuairand1k, S1 and S6 are the domains with the most and least data.
In Fig.~\ref{fig: IA_a}, the S6-specific expert exhibits a lower IA compared to S1, indicating that domains with a smaller data volume are more susceptible to dimensional collapse at the expert level.
In Fig.~\ref{fig: IA_b}, we further categorized items into five groups based on their exposure frequency and observed the collapse of item ID embeddings for both the highest and lowest frequency groups.
We discovered that domain S6 exhibits a lower IA than the overall S6, indicating a more severe collapse, regardless of the group. Within each domain, the low-exposure group consistently shows a lower IA than the high-exposure group. 
We conclude with the following finding:

\begin{tcolorbox}[colback=blue!2!white,leftrule=2.5mm,size=title]
    \emph{%
        Finding 2. Models with domain-specific embeddings suffer from insufficient learning, indicated by dimensional collapse on embeddings of small domains, due to imbalanced data volume across domains.}
\end{tcolorbox}

\begin{figure*}[!h]
\centering
\includegraphics[width=0.9\linewidth]{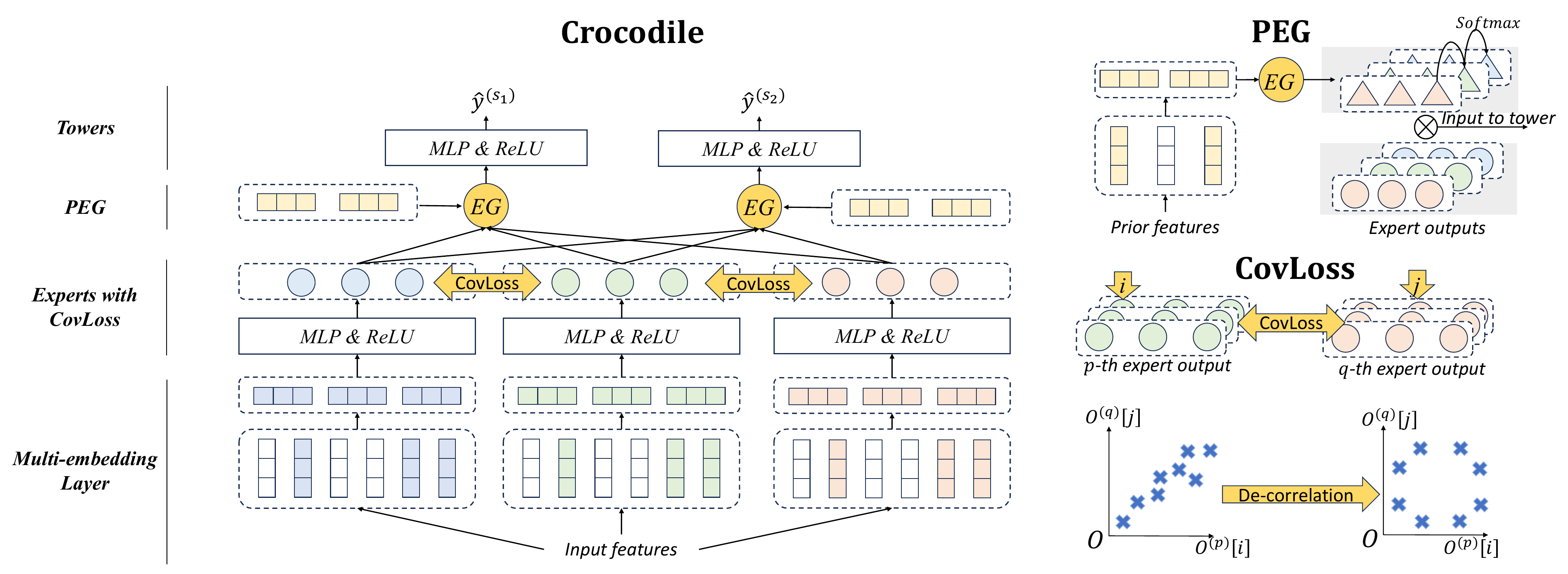}
\caption{Crocodile Architecture, which consists of a Multi-Embedding (ME) layer, a Cross-expert Covariance Loss (CovLoss), and a Prior Informed Element-wise Gating (PEG) mechanism.}
\vspace{-10pt}
\label{fig: crocodile}
\end{figure*}

\section{Our Method: Crocodile}
\label{method}

We propose a multi-embedding architecture with Cross-experts Covariance Loss for Disentangled Learning (Crocodile) to tackle the above-mentioned challenges.
It consists of multiple embeddings and experts, with a disentanglement loss upon them, domain-specific towers, and a customized gating mechanism connecting the experts and the towers, as shown in Fig.~\ref{fig: crocodile}.

\subsection{Multi-Embedding Layer}
The raw input features $x=\{x_1,x_2,\cdots, x_F\}$ contains $F$ fields. 
We employ a multi-embedding (ME) architecture~\cite{su2023stem}, which utilizes $M$ sets of embedding lookup tables. Each embedding looked up from each respective table is fed into the corresponding expert. Thus, in ME architecture, there are $M$ sets of experts.
For a given $p$-th embedding table corresponding to $k$-th sample and $f$-th field, the looked-up embedding is represented as $e_{k, f}^{(p)}$. Thus, each $p$-th embeddings corresponded to $k$-th sample is:
\begin{align}
    e_k^{(p)}= [e_{k, 0}^{(p)}, e_{k, 1}^{(p)}, \cdots, e_{k, F}^{(p)}].
\end{align}

\subsection{Experts and Towers}
Methods based on the Mixture of Experts (MoE) have been widely applied for multi-task~\cite {MMoE2018, PLE2020, su2023stem} or multi-domain~\cite{zhou2023hinet,chang2023pepnet} recommendation. 
Thus, we followed the same architecture and employed several sets of Multi-Layer Perceptrons (MLPs) as the experts and towers. Besides, we utilized a gating mechanism to compute the weights for the representation output from each expert. 
{
The output of the $p$-th expert can be represented as :
\begin{align}
\mathbf{O}_k^{(p)}=MLP^{(p)}(e_k^{(p)}),
\end{align}
where \adjustbox{valign=c}{$\mathbf{O}_k^{(p)}\in \mathcal{R}^{1\times d}$} and ReLU serves as the inter-layer activation function. Through weighted by gating, the outputs \adjustbox{valign=c}{$\mathbf{O}_k^{(\cdot)}$} of each expert are aggregated into a vector, denoted as \adjustbox{valign=c}{$t_k^{(s)}\in \mathcal{R}^{1\times d}$,} the input of $s$-th domain. The detail of gating will be introduced in Section~\ref{gating}.
}

Finally, each domain-specific tower outputs the final predicted value for the corresponding domain as :
\begin{align}
    \hat{y}_k^{(s)} = \sigma(MLP(t_k^{(s)})),
\end{align}
where $\hat{y}_k^{(s)}$ is the predicted value of $k$-th sample in the $s$-th domain, $\sigma(\cdot)$ is the sigmoid function.

\subsection{Cross-expert Covariance Loss}

In order to make the output of experts uncorrelated so as to capture the user's diverse or sometimes even conflicting interests on items, inspired by~\cite{bardes2021vicreg, yu2020learning}, we propose a novel cross-expert covariance loss (CovLoss) to \textbf{explicitly disentangle representations among experts}, defined as:

\begin{align}
\label{covloss}
% \resizebox{0.9\linewidth}{!}{
\mathcal{L}_{Cov}= \frac{1}{d^2}\sum\limits_{p, q \in M\times M, p\geq q}||[\mathbf{O}^{(p)}-\overline{\mathbf{O}}^{(p)}]^T[\mathbf{O}^{(q)}-\overline{\mathbf{O}}^{(q)}]||_1,
% }
\end{align}
where $\mathbf{O}^{(p)}, \mathbf{O}^{(q)}\in \mathcal{R}^{N\times d}$ are the outputs of $p$-th and $q$-th experts,  $\overline{\mathbf{O}}^{(p)}\in \mathcal{R}^{1\times d}$ is the average value of each dimension across all samples, and $||\cdot||_1$ is the $l1$-norm. 

Finally, the overall loss function is defined as :
\begin{align}
\label{overall_formula}
\mathcal{L} = \sum_{s=1}^S [\frac{1}{N_s}\sum_{k=1}^{N_s}\mathcal{L}_{BCE}(\hat{y}_k^{(s)}, y_k^{(s)}) ]+\alpha \mathcal{L}_{Cov},  
\end{align}
where $N_s$ is the number of $s$-th domain's samples, $\hat{y}_k^{(s)}$ is the label of $k$-th sample in domain $s$, and $\alpha$ is the weight of CovLoss.

\subsection{Prior Informed Element-wise Gating Mechanism}
\label{gating}

CovLoss aims to disentangle different representations by minimizing the covariance between dimensions among different experts. However, the existing gating structure limits CovLoss's effectiveness. The gating mechanisms in MMoE~\cite{MMoE2018} and PLE~\cite{PLE2020} assign identical weights to all dimensions within each expert, causing mutual constraints among dimensions. PEPNet's~\cite{chang2023pepnet} Gate Neural Unit operates at the dimension level, but it is designed to scale each dimension separately within the representation rather than to disentangle these representations.

Consequently, we propose an Element-wise Gating mechanism (EG), which implements weight control over corresponding dimensions between experts, as shown in Fig.~\ref{fig: crocodile}:
\begin{align}
    t_k^{s} &= g^{s}_k \odot \mathbf{O}_k,
\end{align}
where $\mathbf{O}_k\in\mathcal{R}^{K\times d}$ is the concatnation across all experts respected to $k$-th sample, and $\odot$ is the element-wise multiplication.

However, when the original ME is deployed in situations with a large number of domains, the scale of parameters becomes unmanageable, because the sets of embeddings are still related to the number of domains.
To address this issue, we proposed Prior Informed Gating (PG), which involves an independent set of domain-unrelated prior embeddings $r_k$ to control the gating.
$r_k$ takes the prior features as input, denoted as $x^{prior}$. For a given $k$-th sample and $f$-th field, the prior embedding is $e_{k, f}^{prior}$ and the input of each gating is: 
\begin{align}
    r_{k} = [e_{k, 0}^{prior}, e_{k, 1}^{prior}, \cdots, e_{k, F'}^{prior}],
\end{align}
where prior embedding contains $F'$ fields. we will select user ID, item ID, and domain ID as the prior embedding fields. 
We defined the PG mechanism as:
\begin{align}
    g^{s}_k &= Softmax(r_k W_g^{s}),
\end{align}
where $r_k\in \mathcal{R}^{1\times l}$ is the prior embedding of $k$-th sample, $W_g^{s}\in \mathcal{R}^{l\times K\times d}$ and $ Softmax(\cdot)$ is operated across the second dimension. 

\subsection{Discussion}
\label{connection}

\paragraph{The Covariance Loss}
Many pioneering work~\cite{shwartz2024information, bardes2021vicreg, ermolov2021whitening} have highlighted the efficiency of tailoring the covariance into objective functions.
Furthermore, some~\cite{hua2021feature} claimed that feature de-correlation by standardizing the covariance matrix can help representations overcome dimensional collapse. 
These efforts have provided evidence to demonstrate the interpretability and efficacy of covariance. 
However, \textbf{they aimed to de-correlate the dimensions within each representation rather than disentangle among representations of all experts}.
Besides, this loss shares the same spirit with the \emph{Maximal Coding Rate Reduction Principle}~\cite{chan2022redunet, yu2020learning} to de-correlate representations from different clusters (experts in our model).

\begin{figure*}[h!]
    \centering
    \includegraphics[width=0.98\linewidth]{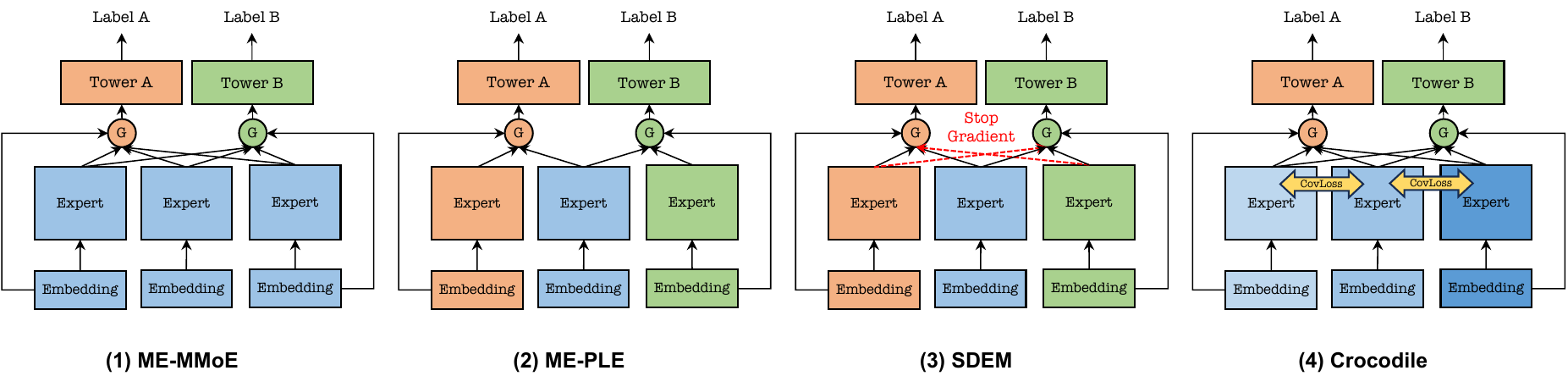}
    \caption{Architecture comparison of different MDL models with multiple embeddings.}
    \label{fig:me_comparison}
\end{figure*}

\paragraph{Connection to Existing Mulit-Embedding MTL or MDL Methods}
The proposed Crocodile differs from those existing multi-embedding-based methods in the following aspects:
1. Different from ME-PLE~\cite{PLE2020,su2023stem} and SDEM~\cite{su2023stem}, which have domain-specific embeddings, all the embeddings and experts in Crocodile and ME-MMoE are shared across domains.
Such design alleviates the insufficient learning of those embeddings corresponding to the small domains as illustrated in Sec.~\ref{subsec:sufficient_learning}.
2. The embeddings and experts in Crocodile are disentangled with the Covariance Loss, making it able to preserve the distinctiveness of each domain, as illustrated in Sec.~\ref{subsec:preserve_distinctiveness}.
Fig.~\ref{fig:me_comparison} shows an architecture comparison of these multi-embedding MDL methods.

\section{Experiment}
\label{experiment}
In this section, we provide comprehensive analyses to address the primary research questions (RQs) outlined below:
\begin{itemize}
    \item RQ1: Can Crocodile outperform state-of-the-art methods on public datasets? Can this be attributed to maintaining domain distinctiveness while enabling sufficient representation learning?
    \item RQ2: Can CovLoss effectively disentangle the expert outputs?
    \item RQ3: How does each proposed component contribute to performance gain?
    \item RQ4: How sensitive is Crocodile to hyper-parameters and model capacity?
    \item RQ5: How does Crocodile perform on a large-scale online recommendation system?
\end{itemize}

The two public datasets correspond to video and product recommendations, respectively. The online revenue of these scenarios is often related to the ranking performance, ensuring that higher-scoring items better match user interests. Hence, we employed two metrics to measure the model performance: AUC and gAUC. AUC measures the overall ranking performance, and gAUC evaluates the quality of intra-user item ranking.

\subsection{Datasets}
In this study, to examine the performance of the model in realistic recommendation, we carefully selected the \textbf{Kuairand1k}~\cite{gao2022kuairand} and \textbf{AliCCP}~\cite{ma2018entire} datasets for our experiments. For Kuairand1k, we chose five domains with top-5 data volumes. Kuairand1k has 30 user-side and 62 item-side features. For AliCCP, we selected all three domains. AliCCP has 12 user-side, 4 item-side, and 4 user-item interaction features. Tab.~\ref{tab: stat} is the statistics for the two datasets after low-frequency filtering. We removed features with fewer than 10 exposures and replaced them with default values. 

Both of these datasets exhibit issues of extreme data imbalance. For instance, in the Kuairand1k dataset, there is a 12-fold difference between the data in S0 and S6. Similarly, in the AliCCP dataset, S2 has 86 times larger data than S3.

\begin{table}[h!]
\caption{Statistics of Kuairand1k and AliCCP datasets.}
\label{tab: stat}
\resizebox{!}{0.1\linewidth}{
\begin{tabular}{@{}lcccc@{}}
\toprule
& \textbf{\#Sample}   & \textbf{\#User}  & \textbf{\#Item}  & \textbf{Domain (\#Impr)}                                                                                     \\ \midrule
Kuairand1k~\cite{gao2022kuairand} & 11,648,743 & 1,000   & 188,148 & \begin{tabular}[c]{@{}l@{}}S0(2.4M)/S1(7.8M)/S2(0.4M)\\ S4(0.9M)/ S6(0.2M)\end{tabular} \\
AliCCP~\cite{ma2018entire}      & 85,316,519 & 241,080 & 499,321 & S1(32M)\,/\,S2(52M)\,/\,S3(0.6M)                                                           \\ \bottomrule
\end{tabular}
}
\vspace{-15pt}
\end{table}

\subsection{Baselines}
To establish a performance benchmark for comparison, we will select comparative methods from both MTL and MDL for multi-domain recommendation. 
We apply the \textbf{SharedBottom}~\cite{SharedBottom1997} method in single/multi-domain contexts and subsequently introduce popular MTL methods such as \textbf{MMoE}~\cite{MMoE2018}, \textbf{PLE}~\cite{PLE2020} (both Single-Embedding and Multi-Embedding~\cite{su2023stem} versions).
Inspired by STEM~\cite{su2023stem}, we also proposed the \textbf{SDEM} method with Shared and Domain-specific EMbeddings. 
Besides, we choose representative MDL methods, including \textbf{STAR}~\cite{sheng2021one}, \textbf{PEPNet}~\cite{chang2023pepnet}, \textbf{HiNet}~\cite{zhou2023hinet}, and \textbf{AdaSparse}~\cite{yang2022adasparse}. 

\subsection{Performance Evaluation (RQ1)}
\label{performance_evaluation}
\begin{table*}[]
\caption{Performance on Kuairand1k dataset. 
Bold and underline highlight the best and second-best results, respectively. * indicates that the performance difference against the
second-best result is statistically significant at \textit{p}-value<0.05.}
\label{tab: kuairand}
\resizebox{!}{0.14\linewidth}{
\begin{tabular}{@{}lc|cccccc|cccccc@{}}
\toprule
\multicolumn{2}{c|}{\multirow{2}{*}{\textbf{Methods}}}                 & \multicolumn{6}{c|}{\textbf{AUC}}                                                                                     & \multicolumn{6}{c}{\textbf{gAUC}}                                                                               \\ \cmidrule(l){3-14} 
\multicolumn{2}{c|}{}                                                  & \textbf{S0}            & \textbf{S1}      & \textbf{S2}      & \textbf{S4}      & \textbf{S6}      & \textbf{Overall} & \textbf{S0}      & \textbf{S1}      & \textbf{S2}      & \textbf{S4}      & \textbf{S6}      & \textbf{Overall} \\ \midrule
\multicolumn{2}{c|}{Single Domain}                                     & 0.73272                & 0.74202          & 0.80179          & 0.72505          & 0.75559          & -                & 0.60336          & 0.59610          & 0.54581          & 0.55085          & 0.60562          & -                \\ \midrule
\multicolumn{1}{l|}{\multirow{3}{*}{MTL (SE) for MDR}} & Shared Bottom & 0.73535                & 0.74271          & \underline{ 0.82033}    & 0.72552          & 0.75945          & 0.78574          & 0.60335          & 0.59694          & 0.56337          & 0.56299          & 0.62500          & 0.66087          \\
\multicolumn{1}{l|}{}                                  & MMoE          & 0.73639                & 0.74279          & \textbf{0.82090} & \underline{ 0.72793}    & 0.76005          & \underline{ 0.78594}    & 0.60188          & 0.59658          & 0.56278          & 0.56325          & 0.62721          & 0.66057          \\
\multicolumn{1}{l|}{}                                  & PLE           & 0.73716                & 0.74271          & 0.81759          & 0.72737          & 0.75807          & 0.78565          & 0.60585          & 0.59710          & 0.56234          & 0.56276          & 0.62908          & 0.66093          \\ \midrule
\multicolumn{1}{l|}{\multirow{3}{*}{MTL (ME) for MDR}} & ME-MMoE       & \underline{ 0.73842}          & \underline{ 0.74281}    & 0.81897          & 0.72637          & 0.75985          & 0.78586          & 0.60649          & \underline{ 0.59874}    & \underline{ 0.56620}    & 0.56080          & 0.63001          & \underline{ 0.66188}    \\
\multicolumn{1}{l|}{}                                  & ME-PLE        & 0.73645                & 0.74240          & 0.80785          & 0.72727          & 0.75138          & 0.78499          & \textbf{0.61084} & 0.59717          & 0.55982          & 0.56170          & 0.61744          & 0.66075          \\
\multicolumn{1}{l|}{}                                  & SDEM       & 0.73741                & 0.74267          & 0.80526          & 0.72776          & 0.75449          & 0.78527          & 0.60820          & 0.59701          & 0.56250          & \underline{ 0.56365}    & 0.62595          & 0.66096          \\ \midrule
\multicolumn{1}{l|}{\multirow{4}{*}{MDL}}              & PEPNET        & 0.73732                & 0.74275          & 0.81396          & 0.72767          & \underline{ 0.76239}    & 0.78574          & 0.60812          & 0.59589          & 0.55435          & 0.55931          & 0.62733          & 0.66018          \\
\multicolumn{1}{l|}{}                                  & HiNet         & 0.73285                & 0.74264          & 0.81955          & 0.72698          & \textbf{0.76451} & 0.78563          & 0.60303          & 0.59830          & 0.56416          & 0.56267          & 0.62790          & 0.66131          \\
\multicolumn{1}{l|}{}                                  & STAR          & 0.71890                & 0.74071          & 0.81586          & 0.72203          & 0.75954          & 0.78336          & 0.58028          & 0.59396          & 0.55229          & 0.55896          & 0.62419          & 0.65652          \\
\multicolumn{1}{l|}{}                                  & AdaSparse     & 0.73445                & 0.74235          & 0.81575          & 0.71845          & 0.76151          & 0.78481          & 0.60653          & 0.59729          & 0.56316          & 0.56128          & \underline{ 0.63150}    & 0.66031          \\ \midrule
\multicolumn{2}{c|}{Crocodile}                                         & \textbf{0.73874} & \textbf{0.74430}* & 0.81646          & \textbf{0.72838} & 0.75817          & \textbf{0.78683}* & \underline{ 0.61075}    & \textbf{0.60068}* & \textbf{0.56964} & \textbf{0.56366} & \textbf{0.63579} & \textbf{0.66373}* \\ \bottomrule
\end{tabular}
}
\end{table*}

\useunder{\uline}{\ul}{}
\begin{table*}[]
\caption{Performance on AliCCP dataset. bold and underline highlight the best and second-best results, respectively. * indicates that the performance difference against the second-best result is statistically significant at \textit{p}-value<0.05.}
\label{tab: aliccp}
\resizebox{!}{0.14\linewidth}{
\begin{tabular}{@{}lc|cccc|cccc@{}}
\toprule
\multicolumn{2}{c|}{\multirow{2}{*}{\textbf{Methods}}}                 & \multicolumn{4}{c|}{\textbf{AUC}}                                         & \multicolumn{4}{c}{\textbf{gAUC}}                                         \\ \cmidrule(l){3-10} 
\multicolumn{2}{c|}{}                                                  & \textbf{S1}      & \textbf{S2}      & \textbf{S3}      & \textbf{Overall} & \textbf{S1}      & \textbf{S2}      & \textbf{S3}      & \textbf{Overall} \\ \midrule
\multicolumn{2}{c|}{Single Domain}                                     & 0.61299          & 0.56544          & 0.61735          & -                & 0.58077          & 0.53023          & 0.58739          & -                \\ \midrule
\multicolumn{1}{l|}{\multirow{3}{*}{MTL (SE) for MDR}} & Shared Bottom & 0.61925          & 0.59355          & 0.61952          & 0.61885          & 0.58908          & 0.57252          & 0.59076          & 0.58982          \\
\multicolumn{1}{l|}{}                                  & MMoE          & 0.61951          & 0.59324          & 0.61969          & 0.61935          & 0.58907          & \textbf{0.57269} & 0.59065          & 0.58992          \\
\multicolumn{1}{l|}{}                                  & PLE           & 0.62152          & 0.59476          & 0.61949          & 0.61931          & 0.58911          & 0.57256          & 0.59070          & 0.58947          \\ \midrule
\multicolumn{1}{l|}{\multirow{3}{*}{MTL (ME) for MDR}} & ME-MMoE       & 0.62271          & {\ul 0.59741}    & {\ul 0.62179}    & 0.62131          & {\ul 0.59002}    & {\ul 0.57258}    & {\ul 0.59148}    & 0.59048          \\
\multicolumn{1}{l|}{}                                  & ME-PLE        & {\ul 0.62352}    & 0.58727          & 0.62063          & 0.62132          & 0.58974          & 0.55777          & 0.59130          & {\ul 0.59049}    \\
\multicolumn{1}{l|}{}                                  & SDEM       & 0.61978          & 0.58187          & 0.61978          & 0.61907          & 0.58941          & 0.55101          & 0.59132          & 0.59036          \\ \midrule
\multicolumn{1}{l|}{\multirow{4}{*}{MDL}}              & PEPNET        & 0.62273          & 0.59239          & 0.61988          & 0.62034          & 0.58851          & 0.56080          & 0.59085          & 0.58975          \\
\multicolumn{1}{l|}{}                                  & HiNet         & 0.62042          & 0.59515          & 0.62003          & 0.62022          & 0.58881          & 0.57191          & 0.59015          & 0.58970          \\
\multicolumn{1}{l|}{}                                  & STAR          & 0.62039          & 0.59204          & 0.61886          & 0.61949          & 0.58705          & 0.56527          & 0.58932          & 0.58855          \\
\multicolumn{1}{l|}{}                                  & AdaSparse     & 0.62192          & 0.59656          & 0.62173          & {\ul 0.62189}    & 0.58819          & 0.57170          & 0.58972          & 0.58924          \\ \midrule
\multicolumn{2}{c|}{Crocodile}                                         & \textbf{0.62485}* & \textbf{0.60033} & \textbf{0.62227} & \textbf{0.62327}* & \textbf{0.59062}* & 0.57237           & \textbf{0.59223}* & \textbf{0.59170}* \\ \bottomrule
\end{tabular}
}
\end{table*}

We compared Crocodile with state-of-the-art recommenders in the \textbf{Kuairand1k}~\cite{gao2022kuairand} and \textbf{AliCCP}~\cite{ma2018entire} datasets. 
All experiments are repeated 6 times and averaged results are reported, as shown in Tab.~\ref{tab: kuairand} and Tab.~\ref{tab: aliccp}, respectively.

\subsubsection{Overall Performance.} 

\textbf{Our proposed Crocodile achieves the best overall AUC and gAUC in two datasets}.
In the Kuairand1k dataset, Crocodile outperforms the second-best method significantly by 0.09\% and 0.19\% in terms of overall AUC and gAUC, respectively. 
In the AliCCP dataset, Crocodile surpasses the second-best method significantly by AUC of 0.138\% and gAUC of 0.12\% than the second-best method. 
Usually, a 0.1\% AUC (gAUC) lift is regarded as a huge improvement in recommendation systems~\cite{zhu2022bars}.

\textbf{Crocodile achieves larger AUC and gAUC almost for almost all domains in two datasets significantly compared to single domain method, with \textit{p}-value<0.05}. 
Such positive transfer can be attributed to Crocodile's preserving domain distinctive and sufficiently representation learning.

\begin{figure}[!h]
\centering
\includegraphics[width=1\linewidth]{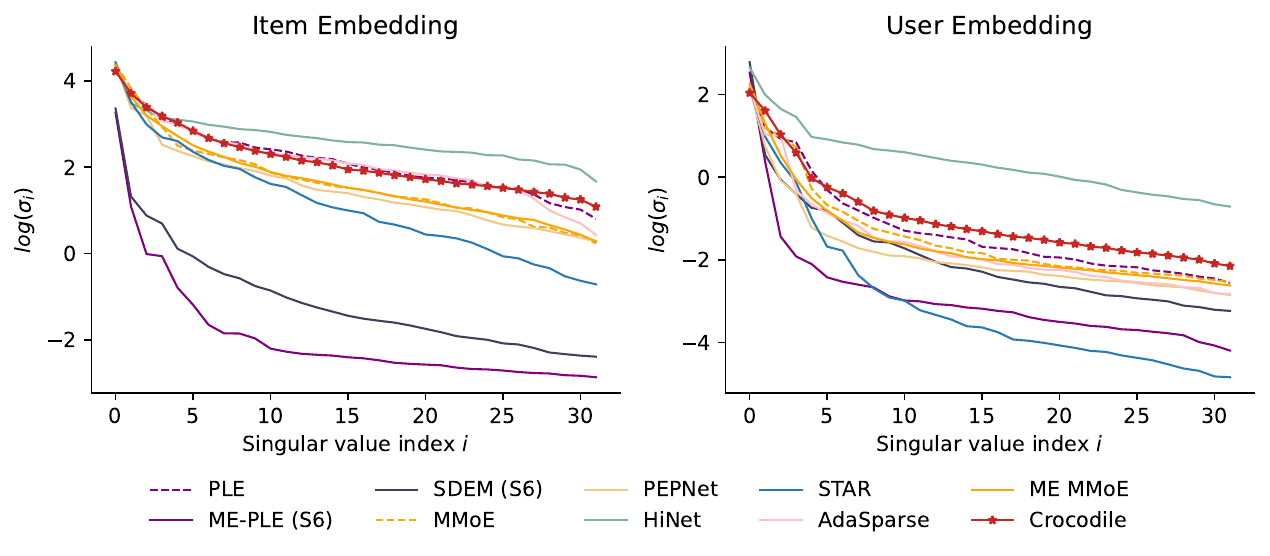}
\caption{Comparison of the singular value spectrum $log(\sigma)$ of item ID and user ID embeddings in Kuairand1k dataset. We reported $log(\sigma)$ of ME-PLE and SDEM S6-specific embedding, while the original $log(\sigma)$ of single embedding or the average of other ME methods.}
\vspace{-20pt} 
\label{fig: collapse}
\end{figure}

\subsubsection{MDL methods with domain-specific embeddings.}
\label{comp_mtl}

The Mulit-Embedding PLE (ME-PLE) and SDEM are two new variants with domain-specific embeddings, inspired by task-specific embeddings in multi-task learning.
However, ME-PLE's overall AUC and gAUC are lower by 0.07\% and 0.02\% compared to PLE, respectively. 
Analyzing each domain, we found the smallest domain experienced the most significant drops in AUC and gAUC, both exceeding 0.3\%. A similar negative transfer was also observed in SDEM. 

To further investigate the mechanism, we examined the quality of representation learning by observing the imbalance in the distribution of singular values~\cite{guo2023embedding,chi2022representation}. As shown in Fig.~\ref{fig: collapse}, We discovered that ME-PLE and SDEM exhibited a clear dominance of top factors with much higher singular values over tail factors in the S6 embeddings. This means that their embeddings of small domains suffer from insufficient learning.
In contrast, Crocodile's embeddings exhibit a more balanced importance of singular values, thereby better supporting the high-dimensional representational space. Consequently, within the smallest domain S6, Crocodile outperforms ME-PLE by 0.68\% in AUC and 1.8\% in gAUC, and surpasses SDEM by 0.37\% in AUC and 0.98\% in gAUC.
Even though SDEM learns the most diverse user interests, it still fails to surpass Crocodile due to insufficient representation learning.

\begin{figure}[h]
\centering
\subfigure[Crocodile vs MDL and MTL] {\includegraphics[width=.235\textwidth]{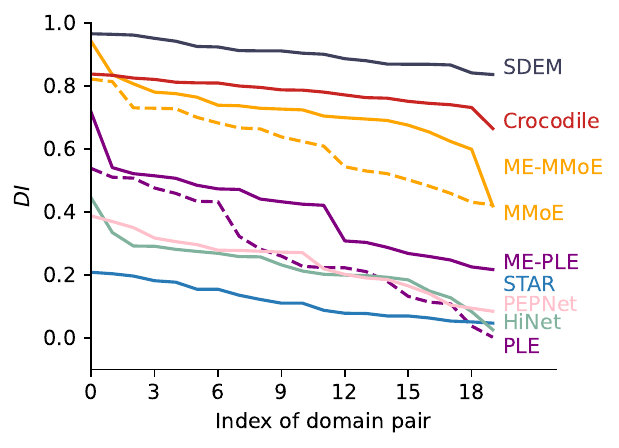}\label{fig: di_a}}
\subfigure[Crocodile vs other Losses] {\includegraphics[width=.235\textwidth]{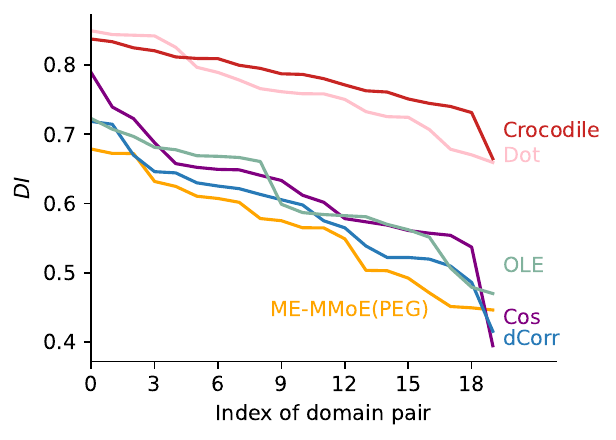}\label{fig: di_b}}
\caption{Diversity Index (DI) of representations in the Kuairand1k dataset. There are five domains in Kuairand1k, resulting in 19 unique combinations. 
We plot the DIs on all combinations in descending order. 
We compared Crocodile with MDL and MTL methods in (a) and other disentanglement losses in (b).}
\label{fig: di}
\vspace{-10pt}
\end{figure}

\begin{figure*}[h!]
\centering
\subfigure[CovLoss between Experts] {\includegraphics[width=.45\textwidth]{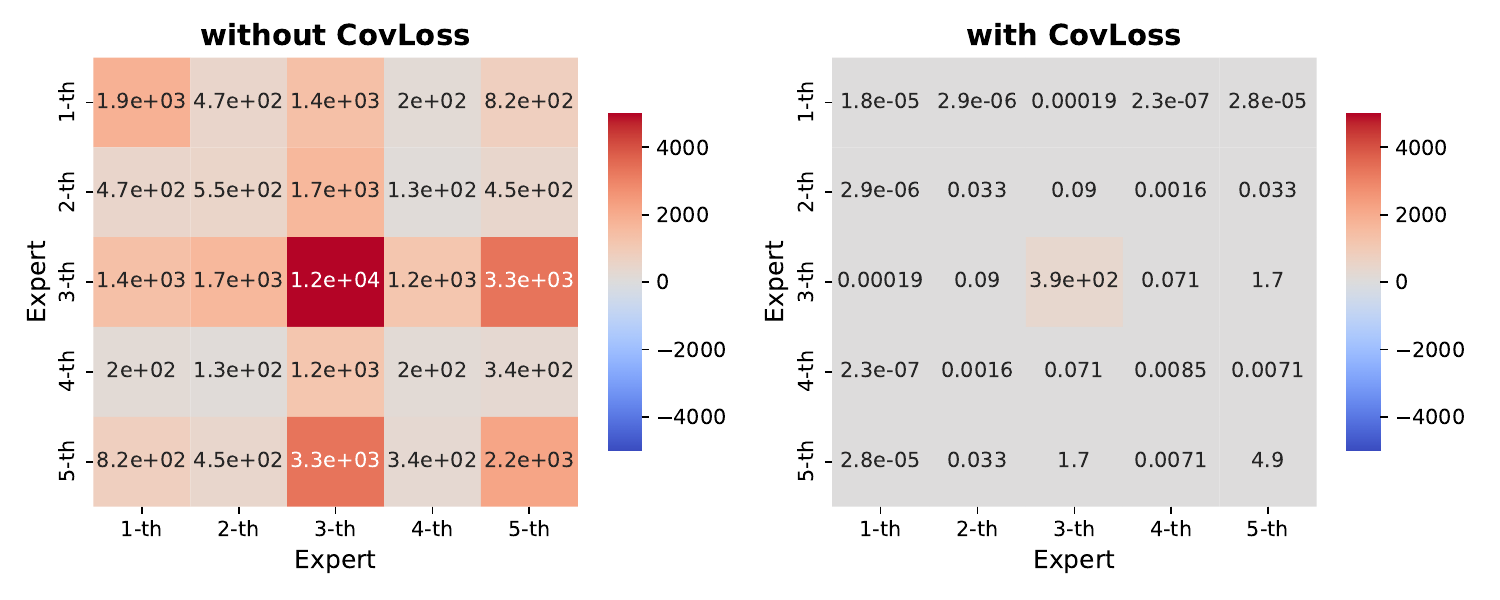}\label{fig: covloss_cmp}}
\subfigure[Covariance across all dimensions of all Experts] {\includegraphics[width=.45\textwidth]{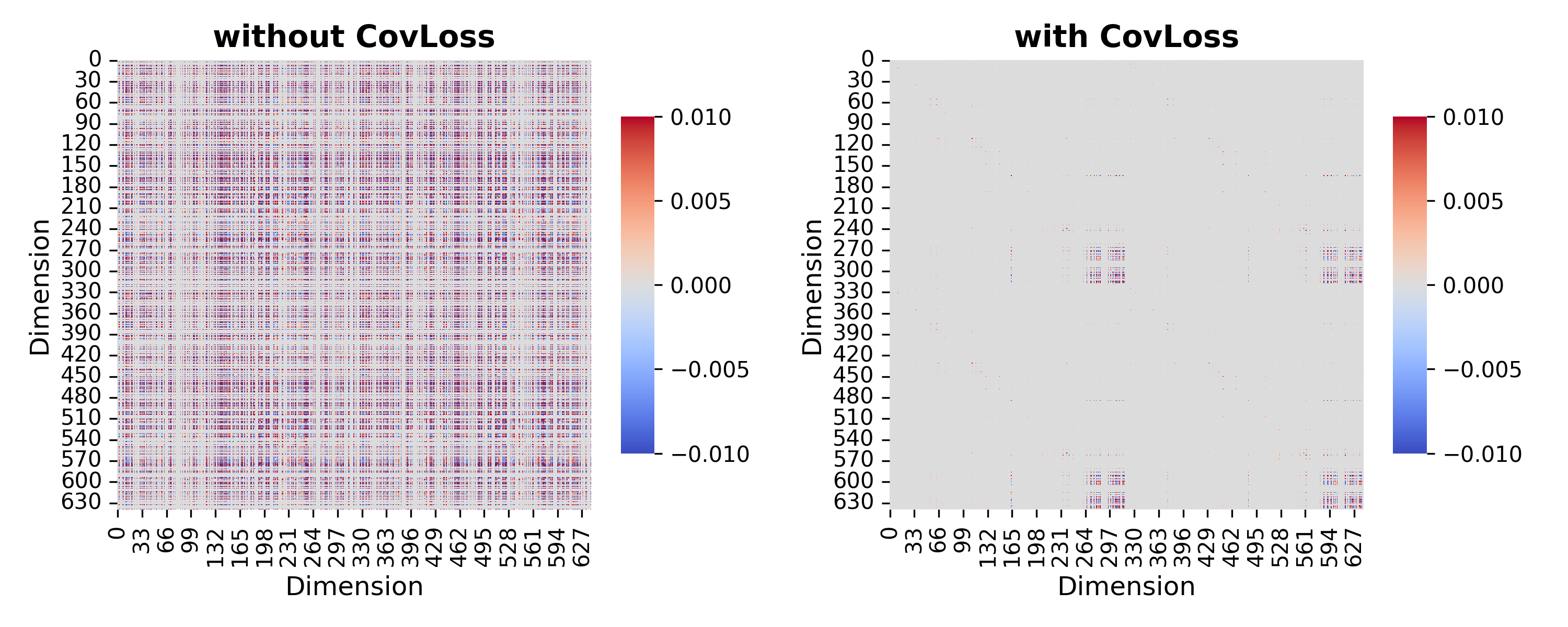}\label{fig: covariance_cmp}}
\caption{Heatmaps of CovLoss between Experts (left) and Covariance across all dimensions of all Experts (right) on Kuairand1k. In the comparison methods with the same structure, "without CovLoss" corresponds to the loss function being only BCE Loss, while "with CovLoss" corresponds to the loss function as the combination of BCE Loss and CovLoss.}
\label{fig: cov_all_compare}
\end{figure*}

\subsubsection{Existing MDL methods.}
\label{comp_mdl}

Multi-domain learning (MDL) methods also exhibit a top-down structural evolution, explicitly capturing domain-specific interests by allocating domain-specific components layer by layer.
We expect that the more thorough the design, the more domain-distinctive the learnable parameters, resulting in better performance. 
We use the IA metric introduced in Sec.~\ref{sec:challenge} to quantify it, as shown in Fig.~\ref{fig: di_a}. Along with the trend from top to bottom, the DI of STAR~\cite{sheng2021one}, HiNet~\cite{zhou2023hinet}, PEPNet~\cite{chang2023pepnet}, and SDEM~\cite{su2023stem} reveals an increasing trend, which is consistent with expectations. However, the performance does not follow this order. 
For example, SDEM has a better DI than HiNet but exhibits lower AUC and gAUC, because performance is also influenced by the sufficiency of representation learning, as discussed in Sec.~\ref{subsec:sufficient_learning}.

In contrast, Crocodile maintains domain distinctiveness more effectively while ensuring adequate representation learning. Although HiNet exhibits more balanced singular values compared to Crocodile, it performs poorly in maintaining domain distinction. Consequently, Crocodile achieves higher AUC and gAUC than HiNet by 0.12\% and 0.24\%, respectively.

\subsubsection{Comparison with Other Forms of Losses.}

In Sec.~\ref{comp_mtl} and ~\ref{comp_mdl}, we found that Crocodile can achieve sufficient representation learning while maintaining domain distinctiveness, resulting in the best performance on two public datasets. This raises the question: can other loss functions achieve similar benefits? We conducted experiments by replacing the loss function in Crocodile, naming the method that retains the structure but uses only BCE Loss as Base.

\begin{table}[h]
\caption{Comparison of different losses with the base model on overall AUC and gAUC in Kuairand1k. * indicates that the performance difference against base is significant at \textit{p}-value<0.05.}
\label{tab: disentangled_loss}
\resizebox{0.85\linewidth}{!}{
\begin{tabular}{@{}l|cc|cc@{}}
\toprule
\multicolumn{1}{c|}{\textbf{Losses}}                                     & \textbf{AUC}     & \textbf{Lift(\%)} & \textbf{gAUC}    & \textbf{Lift(\%)} \\ \midrule
Base                                                                     & 0.78611          & -                 & 0.66208          & -                 \\ \midrule
+Dot~\cite{wu2021vector,tian2023multi}                                                                     & 0.78667          & 0.06\%            & 0.66285          & 0.08\%            \\
+Cos~\cite{liu2021activity}                                                                     & 0.78634          & 0.02\%            & 0.66263          & 0.06\%            \\
+dCorr~\cite{liu2022disentangled}                                                                    & 0.78621          & 0.01\%            & 0.66214          & 0.01\%            \\
+OLE~\cite{lezama2018ole}                                                                     & 0.78650          & 0.04\%            & 0.66319          & 0.11\%            \\ \midrule
+Importance~\cite{shazeer2017outrageously}                                                              & 0.78632          & 0.02\%            & 0.66208         & 0.00\%            \\
+Trans5~\cite{chen2023toward}                                                                  & 0.78550          & -0.06\%           & 0.66017          & -0.19\%           \\ \midrule
\textbf{\begin{tabular}[c]{@{}l@{}}+CovLoss \\ (Crocodile)\end{tabular}} & \textbf{0.78683}* & \textbf{0.07\%}   & \textbf{0.66373}* & \textbf{0.17\%}   \\ \bottomrule
\end{tabular}
}
\vspace{-15pt}
\end{table}

Among the loss functions in Tab.~\ref{tab: disentangled_loss}, dCorr~\cite{liu2022disentangled} shows the smallest improvement, with only a 0.01\% increase in AUC and gAUC, consistent with its poor domain distinctiveness in Fig.~\ref{fig: di_b}. In contrast, \textbf{CovLoss achieves the highest AUC and gAUC improvements of 0.07\% and 0.17\% over the Base model}, highlighting the importance of domain distinctiveness for performance enhancement.

Additionally, we explored other loss functions, such as load balance~\cite{shazeer2017outrageously} and transfer learning~\cite{chen2023toward}, for potential performance gains. As shown in Tab.~\ref{tab: disentangled_loss}, \textbf{Importance} loss~\cite{shazeer2017outrageously} offered limited improvement, while \textbf{Trans5}, an effective transformation loss from \cite{chen2023toward}, had a negative impact. They all fail to surpass CovLoss.

\subsection{CovLoss Analysis (RQ2)}
We are curious about if the CovLoss truly disentangled the experts as expected.
We conducted experiments using the same backbone on Kuairand1k, with and without CovLoss in Fig.~\ref{fig: cov_all_compare}.
We progressively reveal the effectiveness of CovLoss from the course, \textit{i.e.}, expert level, to fine, \textit{i.e.}, dimension level.

Firstly, at the expert level in Fig.~\ref{fig: covloss_cmp}, the method with CovLoss shows significantly lower CovLoss values across each pair of experts compared to the baseline.
For example, the CovLoss between the 3-th and 5-th Experts in the method with CovLoss is $3\times 10^3$ lower than that without CovLoss.

Secondly, at the dimension level in Fig.~\ref{fig: covariance_cmp}, the dimensional-wise heatmap of the method with CovLoss has much smaller number of high-value points than without CovLoss, shows it also has overall significantly smaller covariance compared to the baseline.

Please note that CovLoss acts as an auxiliary loss, serving as a soft constraint and certain learnable parameters must retain intensity information. Thus, few high-value points are still observed in the method with CovLoss.

Finally, in terms of experimental performance, the method with CovLoss achieved AUC and gAUC significant improvements of 0.07\% and 0.17\%, respectively, on the Kuairand1k dataset compared to the method without CovLoss, as shown in Tab.~\ref{tab: ablation}.
These findings demonstrate that CovLoss effectively disentangles the experts and enhances model performance.

\vspace{-5pt}
\subsection{Ablation Study (RQ3)}
\label{ablation}
\begin{table*}[ht!]
\caption{Ablation study in Kuairand1k and AliCCP. ME and SE are shorts for multi- and single-embedding, respectively. B and B+C are shorts for BCE Loss and BCE Loss with CovLoss, respectively. * indicates that the performance difference against Base is significant at \textit{p}-value<0.05.}
\label{tab: ablation}
\resizebox{!}{0.095\linewidth}{
\begin{tabular}{@{}c|ccc|cccc|cccc@{}}
\toprule
\multirow{2}{*}{\textbf{Methods}} & \multicolumn{3}{c|}{\textbf{Components}}             & \multicolumn{4}{c|}{\textbf{Kuairand1k Performance}}                   & \multicolumn{4}{c}{\textbf{AliCCP Performance}}                        \\ \cmidrule(l){2-12} 
                                  & \textbf{Embedding} & \textbf{Gating} & \textbf{Loss} & \textbf{AUC} & \textbf{Lift (\%)} & \textbf{gAUC} & \textbf{Lift (\%)} & \textbf{AUC} & \textbf{Lift (\%)} & \textbf{gAUC} & \textbf{Lift (\%)} \\ \midrule
Base (MMoE)                       & SE                 & -               & B             & 0.78594      & -                  & 0.66057       & -                  & 0.61935      & -                  & 0.58992       & -                  \\ \midrule
\multirow{4}{*}{Ablation Method}  & ME                 & -               & B             & 0.78586      & -0.01\%            & 0.66188       & 0.13\%             & 0.62131      & 0.20\%             & 0.59048       & 0.06\%             \\
                                  & ME                 & PG              & B             & 0.78585      & -0.01\%            & 0.66208       & 0.15\%             & 0.62170      & 0.24\%             & 0.59080       & 0.09\%             \\
                                  & ME                 & PEG             & B             & 0.78611      & 0.02\%             & 0.66208       & 0.15\%             & 0.62227      & 0.29\%             & 0.59117       & 0.12\%             \\
                                  & ME                 & PG              & B+C           & 0.78642      & 0.05\%             & 0.66333       & 0.28\%             & 0.62214      & 0.28\%             & 0.59124       & 0.13\%             \\ \midrule
Crocodile                         & ME                 & PEG             & B+C           & \textbf{0.78683}*      & \textbf{0.09\%}             & \textbf{0.66373}*       & \textbf{0.32\%}             & \textbf{0.62327}*      & \textbf{0.49\%}            & \textbf{0.59170}*       & \textbf{0.18\%}             \\ \bottomrule
\end{tabular}
}
\end{table*}

In this section, we empirically test the effectiveness of our proposed components with an ablation study.
We eliminated CovLoss, PEG, and ME one by one and compared their performance on Kuairand1k and AliCCP, as shown in Tab~\ref{tab: ablation}.

All three components are critical since removing either of them leads to a significant performance drop on both datasets.
In particular, CovLoss is more critical in the Kuairand1k dataset, while CovLoss and EG are equally important in AliCCP.

\vspace{-5pt}
\subsection{Setting Sensitivity Analysis (RQ4)}
\label{rq4}

In this section, we investigate the sensitivity of the model's performance to hyper-parameters, for example, the weight of CovLoss  $\alpha$ and the number of embeddings.
% For hyperparameters, the primary parameter is the weight of CovLoss in Eq.~\ref{overall_formula}, \textit{i.e.,} $\alpha$. 
At Kuairand1k, we find that as long as $\alpha$ is not too small (below $2\times 10^{-5}$), Crocodile's AUC and gAUC remain >0.78590 and >0.66174, thus maintaining its status as the second-best method.

\begin{figure}[!h]
\centering
\includegraphics[width=1\linewidth]{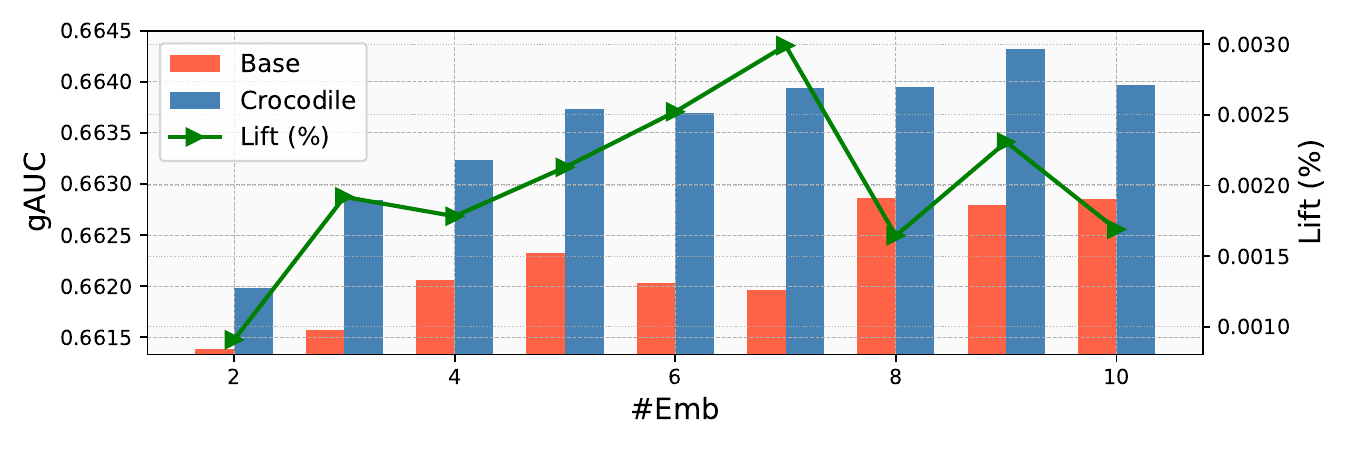}
\caption{gAUC metric of Crocodile varying different numbers of embeddings (\#Emb) compared with Base, \textit{i.e.}, ME-MMoE with PG, in Kuairand1k. The solid line represents the lift ratio.}
\vspace{-10pt} 
\label{fig: num_emb}
\end{figure}

Additionally, we analyzed the impact of model capacity on performance by adjusting the number of embedding sets.
At the minimum number of MEs (\textit{i.e.}, 2-sets), Crocodile achieves a gAUC of 0.66198 and an AUC of 0.78595, which still surpasses other MTL and MDL methods, even when the other MTL(ME) methods employ 5 embeddings sets. 
Moreover, as the number of embeddings increases, the performance gains of Crocodile over ME-MMoE with PG are robust ($+0.06\% \sim +0.20\%$). The detailed results are shown in Fig.~\ref{fig: num_emb}. 

\subsection{Online Deployment (RQ5)}
\label{online}

We conduct an empirical evaluation of Crocodile on Tencent's online advertising platform. 
The underlying model architecture utilizes Heterogeneous Experts with Multi-Embedding framework~\cite{pan2024ad}.
Specifically, our approach involves training multiple feature interaction experts, such as GwPFM~\cite{pan2024ad} (a variant of FFM~\cite{FFM2016} and FwFM~\cite{fwfm2018}), IPNN~\cite{pnn2016}, DCN V2~\cite{wang2021dcn}, or FlatDNN, to capture diverse feature interactions for sparse ID features. 
Additionally, we learn multiple embedding tables for all features~\cite{guo2023embedding, su2023stem, fwfm2018}, with each table corresponding to one or several experts. 
We handle sequence features by TIN~\cite{TIN2024} to capture the semantic-temporal correlation, numeric features by the multiple Numeral Systems Encoding~\cite{N-ary2022, pan2024ad},  and embedding features from external models via Similarity Encoding Embedding~\cite{pan2024ad}.

Specifically, we train a multi-domain model over two main scenarios, with two embedding tables and experts.
The output of these two experts is then fed into the primary scenario.

We conduct online A/B testing on March 2024. 
% baseline
The baseline model employs a common Embedding \& Interaction architecture.
% features
During 5\% A/B test, we observe a 0.11\% AUC lift on several main tasks such as convertible click prediction, and a significant 0.72\% CTR lift and 0.73\% GMV lift.

\paragraph{False Tolerence} 
During online training, if the training samples from the auxiliary domains become unavailable or have some issues, we simply roll back the model to the latest valid checkpoint and keep it trained and updated based on only the main domain until the auxiliary one is recovered.
If the main domain's data pipeline has any problems, we roll back the model and keep it fixed.

\section{Related Work}
\subsection{Multi-domain Recommendation}

Multi-domain learning (MDL) is a promising topic that has garnered attention from large-scale industrial recommendations, such as Alibaba~\cite{sheng2021one,zhang2022leaving,tian2023multi}, Meituan~\cite{chang2023pepnet} and Kuaishou~\cite{zhou2023hinet}. Many MDL methods~\cite{sheng2021one, zhou2023hinet, zhang2022leaving, li2023hamur, chang2023pepnet, tian2023multi, wang2024decoupled} have made great efforts to model complex multi-domain interests. 

Current methods often start from the Shared-bottom~\cite{SharedBottom1997} structure to explore capturing both domain-shared and domain-specific interests. STAR~\cite{sheng2021one} introduces a learnable matrix and dot product operator before outputting the prediction value. HiNet~\cite{zhou2023hinet} and HAMUR~\cite{li2023hamur} allocate domain-shared and domain-specific networks. Furthermore, the PEPNet~\cite{chang2023pepnet} method introduces EPNet, which performs dot product operations between domain-related and domain-shared embeddings, while Maria~\cite{tian2023multi} designs feature scaling, refinement, and correlation modeling to capture cross-domain interests at the feature-level.
They show the trend of integrating domain information at the more bottom level, \emph{i.e., from Tower to Embedding}.

Recently, STEM~\cite{su2023stem} proposed a multi-embedding (ME) structure in multi-task learning (MTL) and step forward in allocating domain-personalized structures. 
It achieved better results compared to both single and multi-embedding versions of MMoE~\cite{MMoE2018} and PLE~\cite{PLE2020} in MTL experiments. 
However, STEM fails in the multi-domain recommendation.
We will discuss the reasons in Sec.~\ref{sec:challenge}.

Besides, some other paradigms are applied for multi-domain recommendation. For example,
AdaSparse~\cite{yang2022adasparse} and ADL~\cite{li2023adl} applied adaptive learning mechanism. 
M2M~\cite{zhang2022leaving} utilizes meta-learning techniques.
EDDA~\cite{ning2023multi}, COAST~\cite{zhao2023cross} and UniCDR\cite{cao2023towards} provided graph learning-based solutions.

\subsection{Disentangled Representation Learning}

Disentangled representation learning helps preserve the distinctiveness of embeddings. Loss-based methods such as OLE~\cite{lezama2018ole}, dot product~\cite{wu2021vector}, and cosine similarity~\cite{liu2021activity} push representations of different domains towards orthogonality and residing in corresponding subspace. Distance correlation (dCorr)~\cite{liu2022disentangled} encourages lower dCorr between different factors. ETL~\cite{chen2023toward} models the joint distribution of user behaviors across domains with equivalent transformation. VAE-based methods~\cite{ma2019learning, wang2022disentangled, liu2022exploiting} promote higher KL divergence between dimensions but often assume a normal prior distribution, which is challenging in the context of sparse interactions~\cite{FM2010}, especially with imbalanced sample volumes in multi-domain recommendation. GAN-based models~\cite{chen2016infogan,yu2021michigan,khrulkov2021disentangled} enhance discrimination among representations but are computationally expensive and unsuitable for online recommendation.

Recently, some research has provided systematic and unified insights. VICReg~\cite{bardes2021vicreg, shwartz2024information} uses covariance and variance as loss components to promote independence across dimensions while preventing insufficient learning. Methods like~\cite{chan2022redunet, yu2020learning} propose a maximal coding rate reduction principle on mixed distributions to maximize the coding rate of the entire representation while minimizing it within each cluster.

\subsection{Dimensional Collapse}

Dimensional collapse~\cite{chi2022representation,guo2023embedding} occurs when all representations span a lower-dimensional space, reflecting the insufficient learning of embeddings. \citet{chi2022representation} found that the SMoE structure~\cite{fedus2022switch,lepikhin2020gshard} may cause dimensional collapse due to gating influences. \cite{guo2023embedding} observed this issue in popular models such as DCN V2~\cite{wang2021dcn} and Final MLP~\cite{mao2023finalmlp}. To measure collapse, they proposed the Information Abundance (IA)~\cite{guo2023embedding, pan2024ad} metric, defined as the ratio between the sum and the maximum \textit{l1-norm} value of the target singular values. A small IA indicates severe insignificance of tail dimensions and potential dimensional collapse.

Some methods~\cite{hua2021feature} bridge the gap between representation disentanglement and dimensional collapse prevention. They also claim that feature decorrelation helps alleviate dimensional collapse by reducing strong correlations between axes, promoting more diverse and informative representations.

\section{Conclusion}
Multi-domain learning is vital in enhancing personalized recommendations. 
Existing methods fail to learn both dimensional-robust representations and diverse user interests.
We propose Crocodile, which employs a \textit{Covariance Loss} (CovLoss) and \textit{Prior informed Element-wise Gating} (PEG) based on the multi-embedding (ME) paradigm.
Both offline and online experiments validate the effectiveness of Crocodile.

\bibliographystyle{ACM-Reference-Format}
\bibliography{7.reference}

\appendix

\section{Computational Complexity}

The Multi-Embedding (ME) paradigm has already been widely adopted in industry~\cite{zhaok2021autoemb, EDDA2023, su2023stem, guo2023embedding, pan2024ad}. 
Compared to these ME architectures, our proposed Crocodile introduces the CovLoss upon it, which only influences the training process while not the inference at all.
\textbf{Its complexity is not only \textit{small and negligible} based on the ME-based backbone}, but it can be further reduced by only computing the loss based on downsampled samples.

The complexity of CovLoss is $\frac{M(M-1)}{2}\times(d^2N)$, shown lower complexity than dCorr~\cite{liu2022disentangled}, which is $\frac{M(M-1)}{2}\times(d^3N+d^2N)$, where $M$ and $N$ are the number of experts and samples, and $d$ is the dimension of expert output. Besides, the complexity of experts in MMoE structure is $M\times\sum_{i=1}^{l-1} Nd_id_{i-1}$, where $d_i$ is the hidden layer dimension for i-th layer in all $l$ layers. This is much higher than CovLoss in practice.

Moreover, we can reduce the complexity of calculating covariance by sampling. We discovered that selecting as low as 32 samples yields no significant difference in computed covariance norms (\textit{p}-value> 0.1) in the testing dataset, contributing as much as 99.2\% complexity reducing, as we originally selected $N=1024$.

\end{document}